\documentclass[twocolumn,aps,superscriptaddress]{revtex4}
\usepackage{amsmath}
\usepackage{graphicx}
\usepackage{tabularx}
\usepackage{amssymb}
\usepackage{hyperref}

\begin{document}

\title{Protecting and accelerating adiabatic passage with time-delayed 
pulse sequences}

\author{Pablo Sampedro}
\affiliation{Departamento de Qu\'imica F\'isica, Universidad Complutense, 28040 Madrid, Spain}

\author{Bo Y. Chang}
\affiliation{School of Chemistry (BK21), Seoul National University, Seoul 151-747, Republic of Korea}

\author{Ignacio R. Sola}
\affiliation{Departamento de Qu\'imica F\'isica, Universidad Complutense, 28040 Madrid, Spain}
\email{isola@quim.ucm.es}

\begin{abstract}
Using numerical simulations of two-photon electronic absorption with
femtosecond pulses in Na$_2$
we show that: i) it is possible to avoid the characteristic saturation or 
dumped Rabi oscillations in the yield of absorption
by time-delaying the laser pulses; ii) it is possible to accelerate
the onset of adiabatic passage by using the vibrational coherence
starting in a wave packet; and iii)
it is possible to prepare the initial wave packet in order to achieve full
state-selective transitions with broadband pulses.
The findings can be used, for instance, to achieve ultrafast 
adiabatic passage by light-induced potentials and understand its
intrinsic robustness.
\end{abstract}


\maketitle

\section{Introduction}

Many technological and chemical processes rely on electronic absorption:
spectroscopic methods, photophysical events, laser-triggered reactions,
to name a few.
The design of linear or nonlinear laser processes that increase the
yield of electronic absorption or make it more robust against
control parameters or experimental uncertainties is a key drive in
the field of Quantum Control\cite{QC1,QC2,QC3,QC4}.
One particular family of methods that has received considerable 
attention is adiabatic passage\cite{STIRAP1,STIRAP2,Shore}.

In adiabatic passage, population inversion between two quantum
states can be achieved independently of the precise peak amplitude
or pulse duration of the field, provided that the {\em pulse area},
that is, the accumulated interaction energy (integrated over time)
is sufficiently large. Typically, adiabatic passage can be achieved
using chirped pulses\cite{ARP1,ARP2,ARP3,ARP4}, 
although transformed-limited ({\it i.e.} Gaussian) pulses can be used 
as well in multiphoton processes.
This is the case in the well known Stimulated Raman Adiabatic
passage or STIRAP scheme\cite{STIRAP1,STIRAP2,Shore}, 
as well as in the strong-field scheme
of Adiabatic Passage by Light-Induced Potentials
or APLIP.\cite{APLIP1,APLIP2,APLIP3,APLIP4,APLIP4b,APLIP5,LAMB1,APLIP6,APLIP7}. 

Both STIRAP and APLIP were proposed (and in the STIRAP case the
scheme has been implemented) using relatively long pulses, such
that a single vibrational state is clearly resolved and selected during
the adiabatic transition. This makes the transition relatively
{\em slow} and the whole scheme more difficult to implement in the
laboratory, as the coherence of the interaction must be preserved
and other possible competing processes have to be
decoupled. This is particularly important when the pulses are strong, 
as in APLIP. APLIP is a very interesting and incredibly robust laser
scheme which, however, demands the use of very strong pulses, such
that other unwanted transitions ({\it e.g.} ionization) usually dominate 
over the selected APLIP process. 
The realization of ultrafast adiabatic passage using
femtosecond pulses would naturally enhance the applicability of the schemes
and the likelihood of implementing APLIP in the laboratory.

However, using femtosecond pulses, such that the process occurs during
the natural time-scale of the vibrational motion, poses other
challenges, that were recently recognized.
%
A typical example is resonant one-photon absorption between electronic states 
of molecules using femtosecond pulses. If the pulses are not too short that 
the nuclear positions are fixed, and not too intense that the potentials are 
completely distorted by the field, the absorption is 
highly suppressed\cite{Par1,Par3}. 
This is a generic behavior of systems with a dense structure of couplings
(that is, without very restrictive selections rules) excited by fields
whose spectral width is larger than the energy spacing of the levels.
Then the initial quantum states are relatively protected from resonant 
transitions from other manifolds. Only very specific states allow population 
transfer avoiding the decoupling between the manifolds, induced by 
Autler-Townes splittings between the unpopulated levels\cite{AT,SS1,SS2}. 
Such states
allow fast parallel transfer of population between the manifolds\cite{Par1,Par3}.
%

\begin{figure}
\includegraphics[width=7cm,angle=-90]{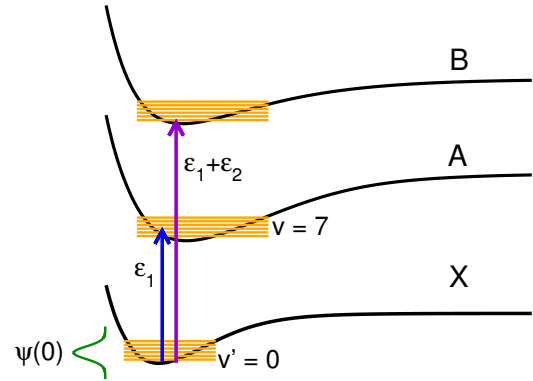}
\caption{Sketch of the model showing the electronic potentials. 
First we study one-photon electronic absorption to the first ($A$) 
electronic band using femtosecond pulses tuned to the $v'=7$ vibrational 
level (Franck-Condon transition). Then we analyze non-resonant two-photon 
absorption to the $B$ band using two pulses tuned to the $v''=0$
vibrational level. The pulses are either blue- or red-detuned from the
$A$ band. We optimize the yield of absorption starting from an initial
wave packet ($\psi(0)$) instead of the ground vibrational level.}
\end{figure}

In this work we  will analyze what happens for non-resonant two-photon
transitions using strong fields, where many assumptions used
to prove the decoupling of the electronic excitation do not operate. 
In particular, the 
potential energy surface of the molecule can be severely deformed
in the presence of the fields, generating light-induced potentials\cite{LIP1,LIP2,LIP3,LIP4}.
Then, the energy eigenstates of the field-free Hamiltonian are not
a very useful basis to follow the dynamics of the system.
Hence, we will study the dynamics in the position representation,
solving the time-dependent Schr\"odinger equation (TDSE) numerically
on a grid, using the Split-operator method\cite{SO}. 
As a particular example, we will analyze one- and two-photon excitation
of the $A$ and $B$ electronic bands in Na$_2$.
The recently developed Geometrical Optimization (GeOp) scheme\cite{Par2} 
will then be used to optimize the initial state, starting in a wave packet, 
in order to maximize the yield of absorption at the lowest possible pulse
intensities, using the initial vibrational coherence to exploit parallel
transfer\cite{Par3}.

Our goal is to understand the necessary conditions for inversion of
the electronic populations (high yields of electronic absorption)
and, secondly, for vibrational state-selective transitions, using
strong pulses of increasingly shorter duration.
In particular we want to understand if the use of time-delayed pulse 
sequences which are needed in adiabatic passage, poses an additional
advantage or disadvantage with respect to the decoupling of electronic
absorption.
%
On the other hand, even for pulses with
bandwidth smaller than the vibrational energy spacing, the population
transfer can occur below the onset of adiabaticity, reducing the
yields of (selective or overall) absorption. 
For such study we will analyze the performance of APLIP with ultrashort pulses,
using the GeOp procedure.


The rest of the manuscript is organized as follows:
In section~2 we will summarize the GeOp procedure applied to our
problem and we will revisit the case of ultrafast electronic
absorption in resonant conditions. The non-resonant case is
analyzed in section~3, where we focus on the decisive role of the pulse 
sequence. In section~4 we study vibrational-state
selective APLIP with ultrashort laser pulses and section~5 is
the conclusions.

\section{Geometrical Optimization}

As a numerical example, we will study one- and two-photon
absorption in Na$_2$, between the $X$, $A$ and $B$ states, shown
in Fig.1. Throughout this work we will assume the Condon limit,
$\mu_{XA}(r) = \mu_{AB}(r) = 1$.
Here we review the application of the GeOp 
procedure\cite{Par2}, applied to wave packet calculations.
Using the Born-Oppenheimer approximation, we will follow the
dynamics of nuclear wave packets $\psi_{X,A,B}$, starting
in $\Psi(0) = (\psi_{X}^{0},0,0)$, the system ground state.
We will use the upper case notation ($\Psi$) for the three-component 
(vector) nuclear wave function. To simplify the notation, whenever 
there is only one component different from zero (or if we are only
interested in a single component) we will use the same subscripts
for the full wave functions as for the components,
$\Psi_X^0 =(\psi_{X}^{0},0,0)$, if no confusion is possible.
Formally, the solution of the TDSE can be written as $\Psi(T) = 
{\sf U}(T,0;{\cal E}(t)) \Psi(0)$, where ${\sf U}$ is the time
evolution operator.
In the simplest GeOp procedure, we maximize the yield of the process 
for a fixed pulse ${\cal E}(t)$ optimizing the initial wave function,
$\Psi^\mathrm{opt}(0) = (\psi_{X}^\mathrm{opt},0,0) = \Psi^\mathrm{opt}_X$ 
with restrictions.
Typically we require it to be a superposition of a small number 
of vibrational eigenstates of the ground electronic state
$\psi_X^\mathrm{opt} = \sum_j^{N_c} a_j \psi_X^j$.
Maximizing the overall yield of absorption $\chi$ to {\it e.g.} the
first electronic band $A$, requires
finding the highest eigenvalue of the {\em reduced yield operator}
${\sf F}^\mathrm{red}$\cite{Par2}, with matrix elements
\begin{equation}
{\sf F}_{jk}^\mathrm{red} = \langle \Psi_X^j | {\sf U}(T,0;{\cal E}(t))
| \Psi_A(T) \rangle \langle \Psi_A(T) | {\sf U}(T,0;{\cal E}(t)) | \Psi_X^k 
\rangle
\end{equation}
Whereas, {\it e.g.} selective excitation of $\psi_B^0$ requires
solving the eigenstates of
\begin{equation}
{\sf F}_{jk}^\mathrm{sel} = \langle \Psi_X^j | {\sf U}(T,0;{\cal E}(t))
| \Psi_B^0 \rangle \langle \Psi_B^0 | {\sf U}(T,0;{\cal E}(t)) | \Psi_X^k 
\rangle
\end{equation}

\begin{figure}
\includegraphics[width=7cm,angle=0]{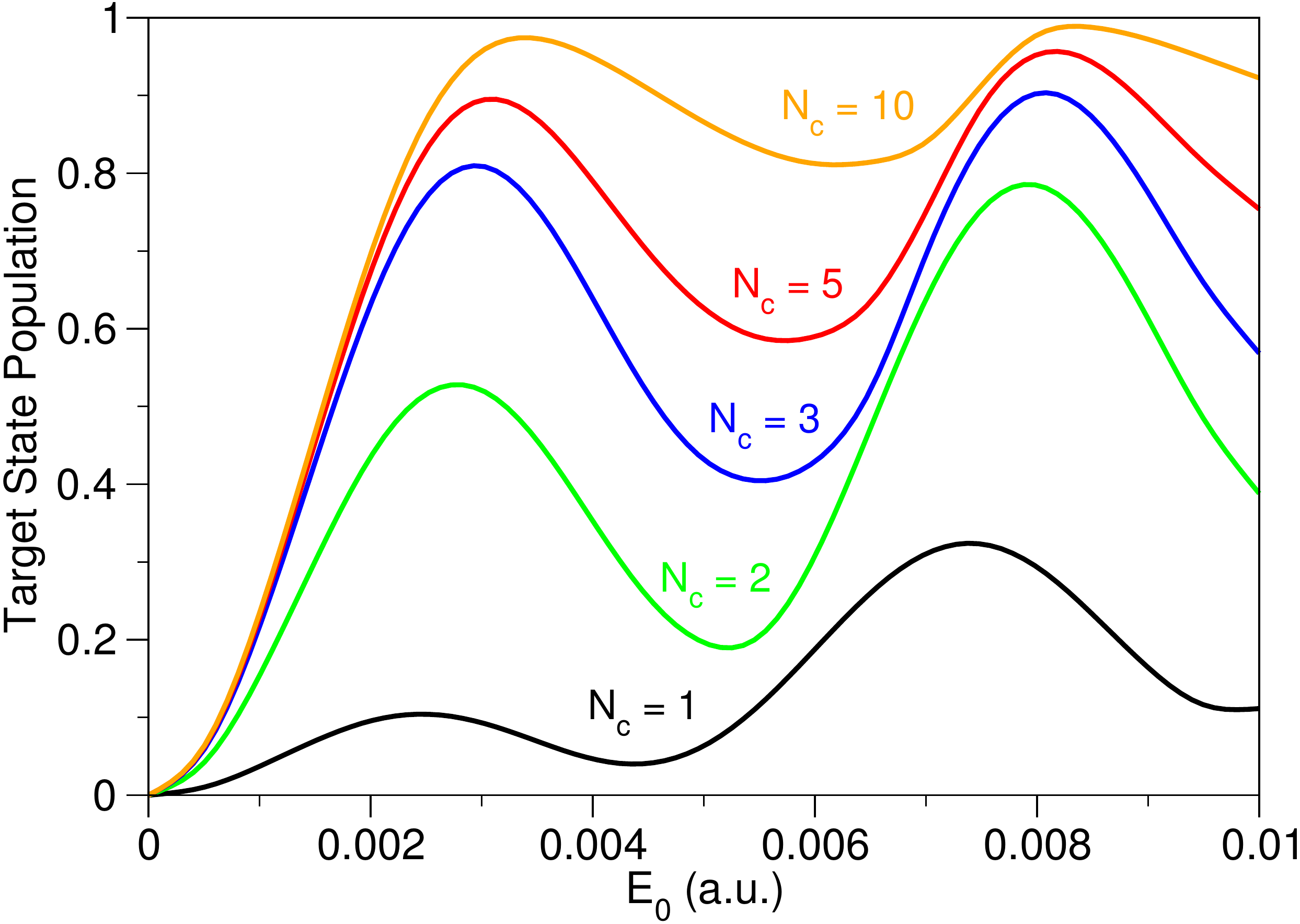}
\caption{Yield of absorption in the $A$ band as a function of the peak
amplitude of the $40$ fs laser pulse. The yield increases when we
start from a wave packet with $N_c$ components (vibrational levels)
instead of a single level. The amplitudes and phases of such wave
packet are found by GeOp.}
\end{figure}

Although we are interested in two-photon processes, for consistency, Fig.2
shows the results of the maximum eigenvalue (optimal yield) for the
electronic absorption to the $A$ state using a $650$ nm $40$ fs 
full-width half-maximum (fwhm) Gaussian pulse as a function of the peak 
amplitude, when the initial state can be a superposition of the first $N_c$ 
vibrational eigenstates of the ground potential. (With $N_c = 1$ the
dynamics is not optimized.)
For the chosen frequency the laser is resonant to the $v=7 \leftarrow v'=0$
transition, but $\sim8$ vibrational levels can participate in the dynamics.
These results are similar to those previously obtained\cite{Par1,Par3} 
but we are now using wave packet calculations and assume the Condon limit.
As expected, the available (within the pulse bandwidth) initially unpopulated 
vibrational states induce the decoupling between the electronic states.
The absorption yield however can be ramped up by optimizing the initial
state using its vibrational coherences, with the yield clearly increasing
with $N_c$. 
Another interesting fact is that the optimization supersedes the pulse area 
theorem\cite{Shore}, 
such that the characteristic Rabi oscillations can be dumped.
Obviously the solution is not robust in the adiabatic sense. For each pulse
intensity a different optimal wave function must be found.

We have shown that for very long pulses the GeOp procedure cannot yield
much better results. On the other hand, for very short pulses GeOp is also
not needed. If the pulse duration $\tau$ is so short that there is no 
vibrational motion involved in the transition, then $\psi_A(\tau) \approx
\psi_X(0)$. Since the $\psi^j_X$ (for different $j$) are orthogonal, 
the matrix ${\sf F}^\mathrm{red}$
is practically diagonal, that is, the initial state (and in fact,
whatever vibrational eigenstate) decouples the Raman transition
and hence avoids the Autler-Townes effect on the electronic absorption.
This is the regime where we can speak of {\em vertical} transitions.

\section{Non-resonant two-photon excitation}

We now turn our attention to non-resonant two-photon excitation of the 
$B$ state. Fig.3 show the results for different pulse sequences using
$40$ fs fwhm Gaussian pulses.
The frequencies of the pulses are chosen
such that the first transition is red-detuned with respect to the resonant
excitation of the $A$ band, using $868$ nm and $710$ nm pulses. The dynamics
is solved using the rotating wave approximation\cite{Shore,LIP4}, 
although for large intensities
the approximation is not really valid. In addition, the probability of
ionization is neglected. Here we are interested in
the qualitative aspects of the process.

Consider first two-photon absorption using coincident pulses, shown
in Fig.3(a).
As in the results for the one-photon resonant transition,
the yield of absorption is dumped by Rabi oscillations,
and it can again be improved by using the vibrational coherences optimizing
the initial state. The main difference with respect to the results shown
in Fig.2 is that the Rabi oscillations are faster, since the effective
coupling (the Rabi frequency) depends with the square of the field amplitude,
that is, with the field intensity.
On the other hand the effect of the vibrational coherence is quite 
noteworthy and optimizing the amplitudes and phases of the first three
vibrational components is typically enough to achieve almost full
population inversion.


\begin{figure}
\includegraphics[width=7cm,angle=0]{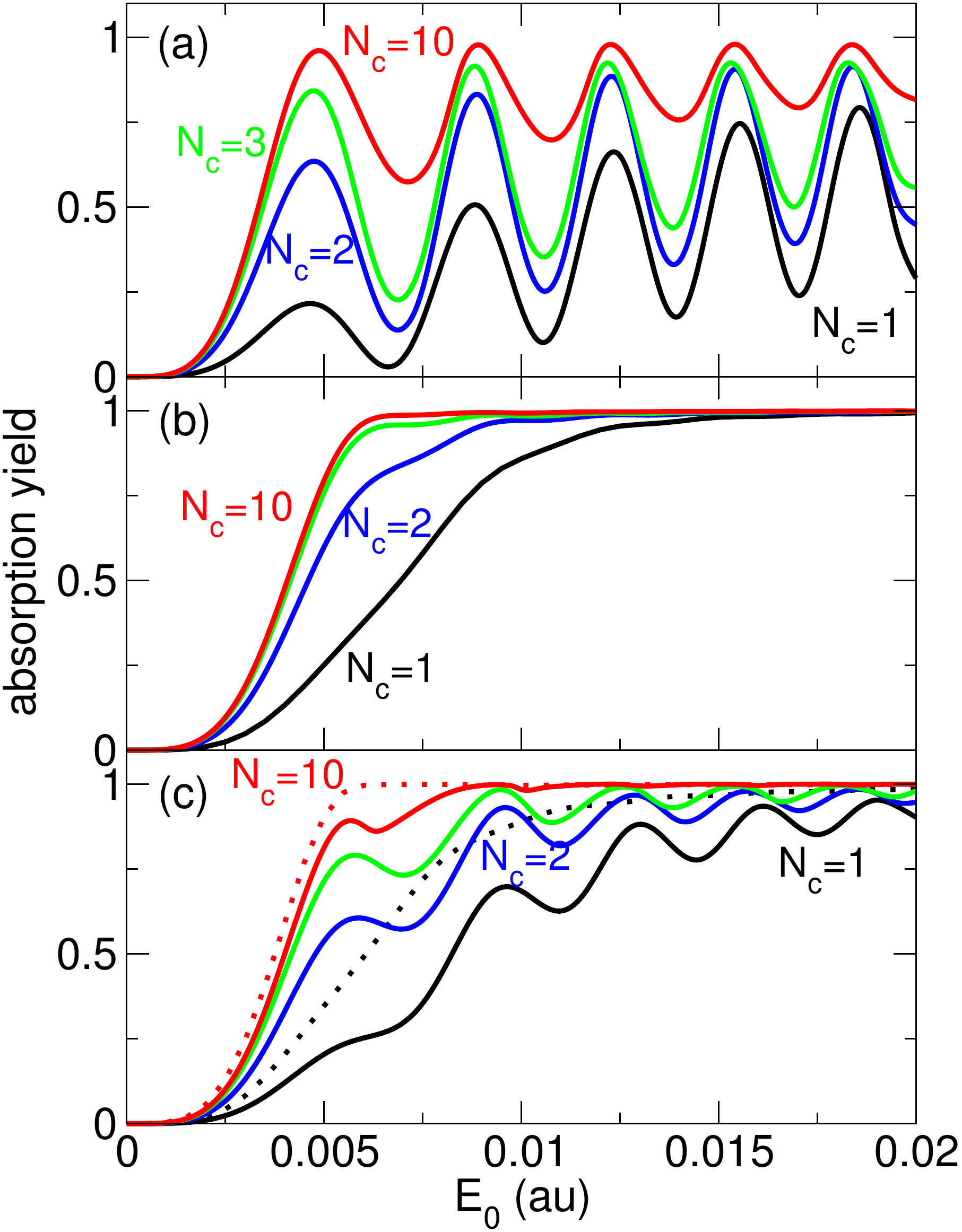}
\caption{Yield of absorption in the $B$ band as a function of the peak
amplitude of the two $40$ fs laser pulses. 
In (a) we use coincident pulses, in (b) and (c) the second pulse is $20$ fs 
delayed with respect to the first pulse. The first pulse is red-detuned 
from the $A$ band to avoid its excitation. In (c) the dotted lines
correspond to choosing $\omega_1$ detuned to the blue of the $A$ band.
The yield increases when the initial state
is an optimized wave packet with $N_c$ components (vibrational levels)
instead of a single level. However, for the time-delayed sequences the
effect is small and mainly reduces the laser-intensity threshold at which 
population inversion occurs.}
\end{figure}

However, the dynamics is completely different when we use time-delayed
pulse sequences. For the results in Fig.3(b) and (c) the time-delay is
set to $20$ fs. In Fig.3(b) the pulse coupling the $X$ and $A$
electronic states, ${\cal E}_1(t)$, precedes the pulse coupling the
$A$ and $B$ states, ${\cal E}_2(t)$, whereas for the results in Fig.3(c)
the opposite order, with ${\cal E}_2(t)$ preceding ${\cal E}_1(t)$ is
chosen. The first pulse sequences is sometimes called in {\em intuitive
order} while the second is considered {\em counter-intuitive}.
Interestingly, the dynamics proceeds now
adiabatically and there are no Rabi oscillations (or very attenuated) in both
pulse sequences. The time-delayed pulses exert an effect similar
to an effective chirp, facilitating the adiabatic passage of the population,
regardless of the dynamic decoupling induced by the initially unpopulated 
levels within the pulse bandwidth. The transition, however, is not
state-selective.
The initial state can be optimized accelerating the onset of full
adiabatic passage, that is, the effective Rabi frequency becomes
larger when the appropriate vibrational coherences are used.

Regarding the order of the sequence, the adiabaticity is better
preserved using the intuitive sequence. However, this is due to the choice 
of pulse frequencies. It is well known that adiabatic passage can proceed 
through a light-induced potential using a counter-intuitive sequence
if the frequencies are chosen such that $\omega_1$ is tuned to the
blue of the first electronic band. This is the basis of the APLIP process.
In Fig.3(c) we show the results for this case (dotted lines), using pulses of 
$566$ and $1259$ nm wavelengths (for $\lambda_1$ and $\lambda_2$, respectively).

\section{Toward ultrafast selective APLIP}

We have shown that 
two-photon transitions with time-delayed pulses and in particular APLIP
are intrinsically protected against Autler-Townes decoupling. We have
optimized the initial state preparing particular superpositions of vibrational
levels to accelerate the onset of the adiabatic passage. However, the
transition is not state-selective. Can we use the GeOp procedure to
achieve ultrafast selective APLIP?

\begin{figure}
\includegraphics[width=7cm,angle=0]{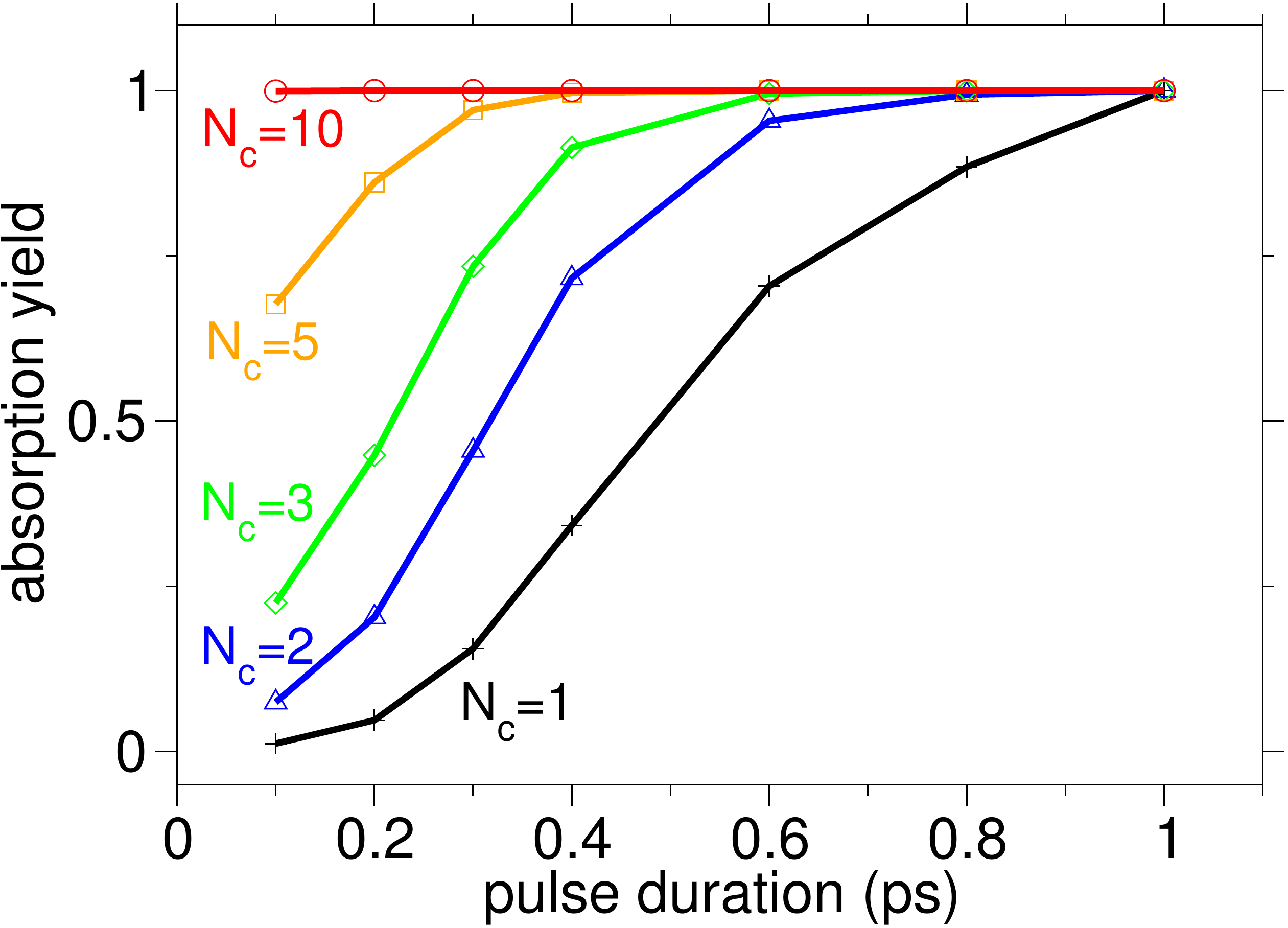}
\caption{Yield of selective excitation of the ground vibrational
level of the $B$ state, as a function of the pulse duration (fwhm), using
a counterintuitive pulse sequence with peak amplitude of ${\cal E}_0
= 0.01$ a.u.}
\end{figure}

In Fig.4 we show the yield of selective excitation of the $\Psi_B^0$
state using the GeOp method [Eq.(2)] with pulses of increasing duration 
using the APLIP pulse sequence with peak field amplitudes ${\cal E}_0 =
0.01$ a.u. and the previous wavelengths. The time-delay is set as half
the fwhm. Using pulses longer than $1$ ps the APLIP scheme is selective.
Although for all cases considered here there is electronic population
inversion, excited vibrational levels are increasingly populated using
shorter pulses. For durations of $100$ fs the
selective yield is practically zero. However, starting in optimized
superpositions states one can recover the full state-selective population
transfer. The number of vibrational levels of the $X$ electronic
state that must be included in the optimal initial wave function must
increase as the pulses become shorter.

\begin{figure}
\includegraphics[width=7cm,angle=0]{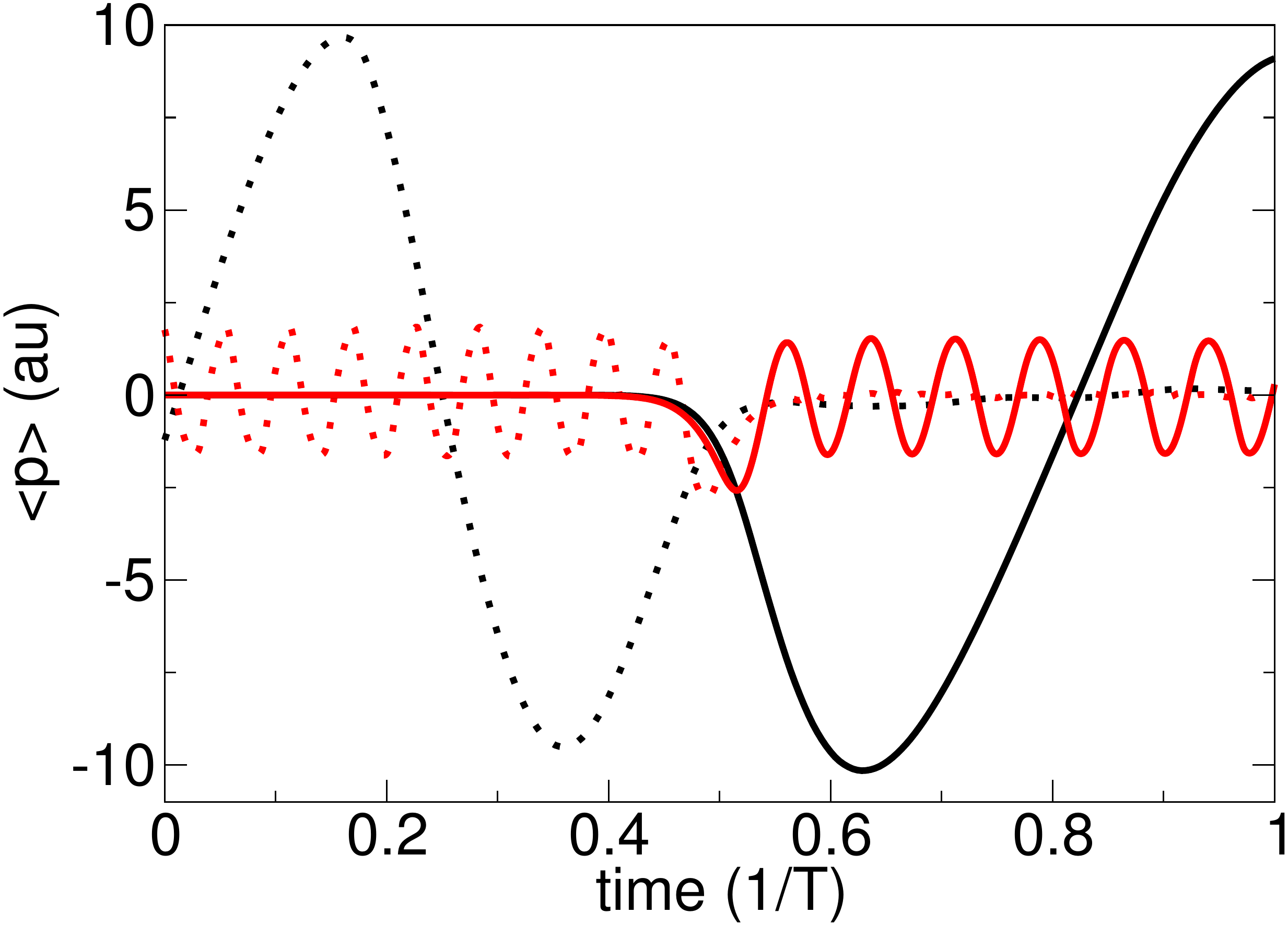}
\caption{Average momentum of the wave packet without optimizing
the initial state (dashed lines) and for the optimal initial
wave packet (full line) with APLIP using long $800$ fs (red)
or short $100$ fs (black) pulses. The time axis is scaled with
respect to the total time propagation $T$.}
\end{figure}

In order to understand the optimization mechanism one must consider the
source of adiabaticity in the transfer. In APLIP the initial wave
function is propagated on a light-induced potential (LIP) that connects
the ground state with the second excited electronic state. During
the process the wave packet must remain in the bottom of the LIP, that
is, it must be the ground vibrational instantaneous eigenstate of the
LIP. This requires the process to be quasi-static. Changes in the LIP
must be done so slow as to allow the wave function to adapt, in a 
Born-Oppenheimer way. If the pulses are short, the changes in 
the LIP impart a momentum on the wave function inducing vibrational excitation.
In Fig.5 we show the evolution of the average momentum of the system
wave function $\langle \Psi(t) | p | \Psi \rangle \equiv \langle p(t) \rangle$. 
The selective transfer can only be regained when the initial wave packet 
has an initial (opposite) momentum $\langle p(0) \rangle$ that compensates 
the effect of the field on the wave function.

\begin{figure}
\includegraphics[width=7cm,angle=0]{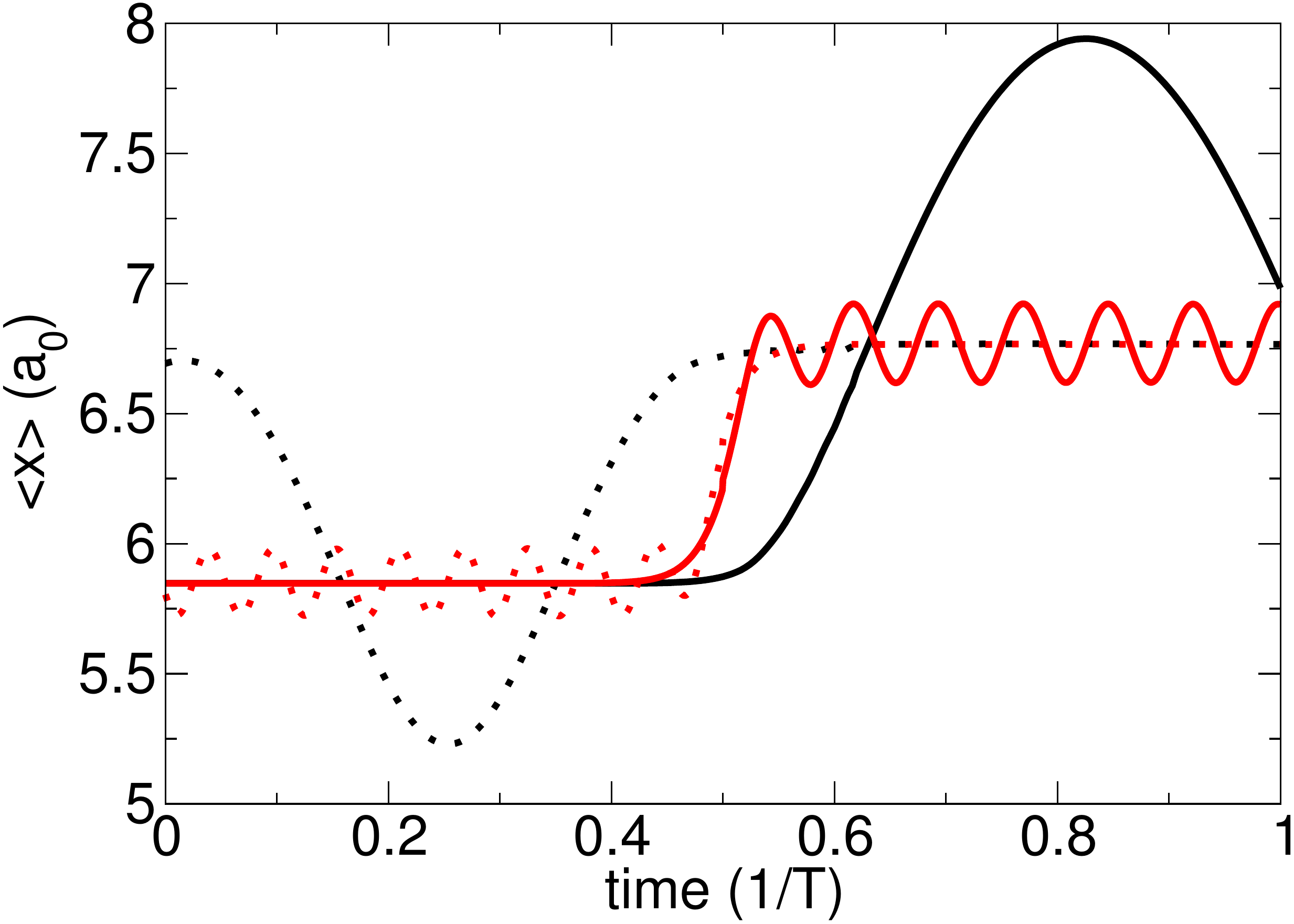}
\caption{Average position of the wave packet without optimizing
the initial state (dashed lines) and for the optimal initial
wave packet (full line) with APLIP using long $800$ fs (red)
or short $100$ fs (black) pulses. The time axis is scaled with
respect to the total time propagation $T$.}
\end{figure}

The time-symmetry that is apparent in Fig.5 can also be observed in the
dynamics in the position representation shown in Fig.6, where we
use $100$ fs (fwhm) pulses with ${\cal E}_0 = 0.01$ a.u.
Allowing for enough vibrational states in the initial optimized wave
function, the GeOp procedure essentially provides the same initial
solution as that obtained by propagating backwards in time the target
wave function, $\Psi_X^\mathrm{opt} \approx {\sf U}(0,T,{\cal E}(t)) |\Psi_B^0
\rangle$.

\section{Conclusions}


The vibrational coherence plays an important role in accelerating
electronic absorption. Hence, preparing an initial wave packet opens
the gate to new schemes of ultrafast population inversion\cite{IRcontrol1,IRcontrol2,IRcontrol3,IRcontrol4}.
In this work we use the recently proposed GeOp procedure 
to obtain the optimal parameters of the initial wave packet.
This is essential in resonant one-photon transitions, as the Autler-Townes
splittings considerably reduce the yield of absorption. 
However, by using time-delayed pulse sequences the yield of non-resonant
two-photon absorption is already protected from the Autler-Townes 
splittings. 
Time-delayed pulse sequences possess an inherent robustness
that make two-photon transitions particularly interesting for quantum
control of population transfer.
However, an initial wave packet can still be used to reduce
the onset of adiabatic passage, making population inversion possible
with weaker or shorter fields, which pose additional advantages to
the experimental implementation of such schemes.
This is particularly the case in APLIP. We have shown that one
can prepare ultrafast-APLIP scenarios with ultrashort pulses and moreover, 
the APLIP process can be state-selective provided that the initial wave
packet overlaps a sufficiently large number of vibrational levels.

\section*{Acknowledgment}
We thank St\'ephane Gu\'erin for his stimulating questions motivating this work.
The work was supported by the International cooperation program 
(NRF-2013K2A1A2054518) and the Basic Science Research program 
(NRF-2013R1A1A2061898) funded by the Korean government,
and the Spanish MICINN project CTQ2012-36184.
I. R. S. acknowledges support from the Korean Brain Pool Program.


\begin{thebibliography}{10}


\bibitem{QC1}
Rice S A and Zhao M 2000 {\it Optical Control of Molecular Dynamics} 
(New York:John Wiley \& Sons)
\bibitem{QC2}
Shapiro M and Brumer P 2012 {\it Quantum Control of Molecular Processes, 2nd,
Revised and Enlarged Edition}
(Weinheim:Wiley-VCH)
\bibitem{QC3}
D'Alessandro D 2008 {\it Introduction to Quantum Control and Dynamics}
(Boca Raton:Chapman \& Hall)
\bibitem{QC4}
Brif C, Chakrabarti R and Rabitz H 2012 {\it Adv. Chem. Phys.} 148 1-76.

\bibitem{STIRAP1}
Bergmann K, Theuer H and Shore B W 1998 {\it Rev. Mod. Phys.} 70 1003
\bibitem{STIRAP2}
Vitanov N V, Halfmann T and Shore B W, Bergmann K 2001
{\it Annu. Rev. Phys. Chem.} 52 763-809
\bibitem{Shore}
Shore B W 2011 {\it Manipulating Quantum Structures Using Laser Pulses}(Cambridge:Cambridge University Press)

\bibitem{ARP1}
Melinger J S, Gandhi S R, Hariharan A, Tull J X and Warren W S 1992
{\it Phys. Rev. Lett.} 68 2000
\bibitem{ARP2}
Band Y B and Magnes O 1994 {\it Phys. Rev. A} 50 584
\bibitem{ARP3}
Cao J, Bardeen C J and Wilson K R 1998 {\it Phys. Rev. Lett.} 80 1406
\bibitem{ARP4}
Malinovsky V S and Krause J L 2001 {\it Eur. Phys. J. D} 14 14

\bibitem{APLIP1}
Garraway B and Suominen K -A, 1998 {\it Phys. Rev. Lett.} 80 932
\bibitem{APLIP2}
Rodriguez M, Suominen K -A and Garraway B 2000 {\it Phys. Rev. A} 62 053413
\bibitem{APLIP3}
Sola I R, Santamaria J and Malinovsky V 2000 {\it Phys. Rev. A} 61 043413
\bibitem{APLIP4}
Sola I R, Chang B Y, Santamaria J, Malinovsky V and Krause J L 2000 {\it J, Phys. Rev. Lett.} 85 4241
\bibitem{APLIP4b}
Chang B Y, Sola I R, Santamaria J, Malinovsky V S and Krause J L 2001
{\it J. Chem. Phys.} 114 8820
\bibitem{APLIP5}
Malinovsky V S, Santamaria J and Sola I R 2003 {\it J. Phys. Chem. A}
107 8259
\bibitem{LAMB1}
Chang B Y, Rabitz H and Sola I R 2003 {\it  Phys. Rev. A} 68 031402
\bibitem{APLIP6}
Gonz\'alez-V\'azquez J, Sola I R and Santamaria J 2006 {\it J. Phys. Chem. A}
110 1586
\bibitem{APLIP7}
Suominen K -A 2014 {\it J. Mod. Opt.} 61 851


\bibitem{Par1}
Chang B Y, Shin S and Sola I R 2015 {\it J. Phys. Chem. Lett.} 6 1724
\bibitem{Par3}
Chang B Y, Shin S and Sola I R 2015 {\it J. Phys. Chem. A} 119 9091

\bibitem{AT}
Autler S H and Townes C H 1955 {\it Phys. Rev.} 100 703
\bibitem{SS1}
Townsend D, Sussman B J and Stolow A 2011 {\it J. Phys. Chem. A} 115 357-373
\bibitem{SS2}
Sola I R, Gonz\'alez-V\'azquez J, de Nalda R and Ba\~nares L 2015
{\it Phys. Chem. Chem. Phys.} 17 13163-13772

\bibitem{LIP1}
Yuan J M and George T F 1978 {\it J. Chem. Phys.} 68 3040
\bibitem{LIP2}
Bandrauk A D and Sink M L 1981 {\it J. Chem. Phys.} 74 1110
\bibitem{LIP3}
Bandrauk A D and Aubanel E E and Gauthier J M 1994 {\it Molecules in laser field, Chapter 3} ed A D Bandrauk (New York:Dekker)
\bibitem{LIP4}
Chang B Y, Sola I R and Shin S 2016 {\it International Journal of Quantum Chemistry} DOI: 10.1002/qua.25066

\bibitem{SO}
Feit M, J. Fleck Jr J and Steiger A 1982
{\it J. Comp. Phys.} 47 412–433

\bibitem{Par2}
Chang B Y, Shin S and Sola I R 2015 {\it J. Chem. Theor. Comput.}
11 4005

\bibitem{IRcontrol1}
Amstrup B and Henriksen N E 1992
{\it J. Chem. Phys.} 97 8285-8295
\bibitem{IRcontrol2}
Meyer S and Engel V 1997
{\it J. Phys. Chem. A} 101 7749-7753
\bibitem{IRcontrol3}
Elghobashi N, Krause P, Manz J and Oppel M 2003 {\it Phys. Chem. Chem. Phys.}
5 4806-4813
\bibitem{IRcontrol4}
Elghobashi N and Gonz\'alez L 2004
{\it Phys. Chem. Chem. Phys.} 6 4071-4073



\end{thebibliography}
\end{document}